\documentclass[12pt]{article}
\usepackage{amsmath}
\usepackage{graphics}
\usepackage{comment}

\usepackage{natbib}

\usepackage{bibentry}
\usepackage{subcaption}
\usepackage{diagbox}
\usepackage{enumerate}

\usepackage{tikz}
\usetikzlibrary{shapes, arrows}
\usepackage{multirow}
\usepackage{microtype}
\usepackage{graphicx}
\usepackage{subcaption}
\usepackage{booktabs}

\usepackage{amssymb}
\usepackage{mathtools}
\usepackage{amsthm}

\usepackage{float}
\theoremstyle{plain}
\newtheorem{theorem}{Theorem}[section]

\theoremstyle{definition}
\newtheorem{definition}[theorem]{Definition}

\theoremstyle{remark}

\usepackage{bibentry}
\usepackage{color}

\usepackage{mathrsfs}

\graphicspath{{./art/}}

\usepackage[plain,noend]{algorithm2e}

\makeatletter
\renewcommand{\algocf@captiontext}[2]{#1\algocf@typo. \AlCapFnt{}#2} 
\def\@algocf@capt@plain{top}
\renewcommand{\algocf@makecaption}[2]{%
  \addtolength{\hsize}{\algomargin}%
  \sbox\@tempboxa{\algocf@captiontext{#1}{#2}}%
  \ifdim\wd\@tempboxa >\hsize
    \hskip .5\algomargin%
    \parbox[t]{\hsize}{\algocf@captiontext{#1}{#2}}
  \else%
    \global\@minipagefalse%
    \hbox to\hsize{\box\@tempboxa}
  \fi%
  \addtolength{\hsize}{-\algomargin}%
}
\makeatother

\begin{document}

%
%

\title{\bf Demographic Parity-aware Individualized Treatment Rules }

%
%
%
%
%

\author{Wenhai Cui\\
    Department of Applied Mathematics, The Hong Kong Polytechnic University,\\
    Wen Su \\
    Department of Biostatistics, City University of Hong Kong,\\
    Xiaodong Yan\\
    Department of Mathematics and Statistics, Xi'an Jiaotong University,\\
    Donglin Zeng\\
    Department of Biostatistics, University of Michigan,\\
    Xingqiu Zhao\\
    Department of Applied Mathematics, The Hong Kong Polytechnic University.\\
   }

\maketitle

\begin{abstract}
There has been growing interest in developing optimal individualized treatment rules (ITRs) in various fields, such as precision medicine, business decision-making, and social welfare distribution. The application of ITRs within a societal context raises substantial concerns regarding potential discrimination over sensitive attributes such as age, gender, or race.
To address this concern directly, we introduce the concept of demographic parity  in ITRs. However, estimating an optimal ITR that satisfies the demographic parity  requires solving a non-convex constrained optimization problem. To overcome these computational challenges, we employ tailored fairness proxies inspired by demographic parity and transform it into a convex quadratic programming problem. Additionally, we establish the consistency  of the proposed estimator and the risk bound. The performance of the proposed method is demonstrated through extensive simulation studies and real data analysis.

\end{abstract}


\vfill

\section{Introduction}
\label{sec:intro}

Incorporating individual characteristics into the decision-making process offers significant potential for improving efficiency and precision. Such practice has become prevalent in various fields, such as precision medicine \citep{shi2018high, wang2018learning}, social welfare distribution \citep{rosholm2010reducing}, ridesharing \citep{shi2022off}, and mobile health \citep{shi2022statistically}.
The three main approaches for estimating individual treatment rules (ITR) include Q-learning \citep{dayan1992q,chakraborty2010inference, qian2011performance, song2015penalized}, A-learning \citep{robins2000marginal,murphy2003optimal,robins2004optimal}, and model-free policy search methods such as outcome weighted learning (OWL) \citep{zhao2012estimating} as well as its extended doubly robust version \citep{liu2018augmented}.

The optimal ITR is commonly estimated by maximizing its value function. However, implementing such an optimal ITR directly in our society may raise concerns about fairness. Discrimination based on sensitive attributes such as nationality, religion, income, race, or gender can result in unjust treatment outcomes. Furthermore, subjective variables like interviewer scores can be influenced by these attributes. Estimating the optimal ITR using biased data can damage public trust and contradict policymakers' original intentions.
For instance, in social welfare distribution,  eligibility paradoxes can arise, where individuals in genuine need may be excluded due to gender or racial biases, while those who receive assistance may not actually require it \citep{heidari2018fairness}. Similarly, in healthcare, older women were less likely to be admitted to the intensive care unit and more likely to die following critical illness than men \citep{fowler2007sex,penner2010aversive,williams2015racial}.
Furthermore, such unfairness can also be observed in the domain of precision advertising recommendations, where algorithms tend to disproportionately present science and engineering job advertisements to male audiences, thereby unintentionally marginalizing qualified female candidates \citep{lambrecht2019algorithmic}.
 Considering these concerns, it is crucial to develop a policy that strikes a balance between fairness and efficiency when considering individual characteristics for decision-making.

Most existing approaches for estimating optimal ITR do not consider fairness \citep{zhao2012estimating,shi2018high,shi2016robust}.
While traditional methods may remove sensitive attributes from the training data, this does not ensure fairness.  Other variables could still be correlated with these sensitive attributes, causing biased outcomes \citep{penner2010aversive,williams2015racial}. As shown in Experiment 1 in Section 5.1, individuals with certain sensitive attributes may have a significantly reduced chance of receiving welfare benefits compared to those without these attributes, even when the optimal ITR is estimated without considering these sensitive attributes.

The fairness aspect of the Individualized Treatment Regime (ITR) has been relatively understudied in the literature. \cite{fang2022fairness} introduced a fairness-oriented policy framework that relies on quantile constraints. Additionally, \cite{viviano2023fair} utilized the Pareto principle to establish fairness rules, primarily focusing on linear policies. Another study by \cite{kim2023fair} examined the difference in the conditional average treatment effect between two observed values of a one-dimensional sensitive attribute. However, there is a scarcity of research focusing on the independence between the ITR and sensitive attributes with multiple dimensions. This independence is crucial for decision-makers to ensure a fair distribution of welfare benefits across various subgroups.

In this study, we introduce the concept of ``Demographic Parity-aware Individualized Treatment Rule", which aims to achieve demographic balance in policies.
To address the challenges of estimating Demographic Parity ITR in datasets with inherent biases, particularly the non-convex nature of Demographic Parity constraints, we employ a two-step approach. First, we identify a Demographic Parity surrogate, and then we employ Lagrangian dual techniques to convert the initial problem into a solvable convex quadratic programming problem. The main goal of our approach is to maximize the overall treatment benefit while adhering to the principles of Demographic Parity.

Our main contributions are as follows:
\begin{itemize}

\item[(i)]    We propose an innovative mathematical framework for fair ITR, designed to achieve demographic parity. Additionally, we introduce an adjustable  mechanism that allows policymakers to set different levels of fairness for the estimated optimal ITR.

\item[(ii)] The proposed fairness proxy effectively captures discrimination by considering the complex  non-linear correlations between specific demographic factors and the decision function. This  proxy significantly  enhances fairness in decision-making processes.

\item[(iii)] The algorithm developed in this study formulates the estimation of the optimal fair ITR as a convex quadratic programming problem, which can be solved efficiently. Additionally, kernel techniques are employed to  handle scenarios where the data are not linearly separable, further improving the algorithm's performance.
\end{itemize}

\section{ {Preliminaries}}

\subsection{Individualized treatment rules}
In a randomized trial, for each individual, we observe the tuple $(\boldsymbol{X}, \boldsymbol{S}, A, R)$, where $\boldsymbol{X}=(X_1,  \ldots, X_p)^\top$ is a $p$-dimensional vector of covariate,  $\boldsymbol{S}=(S_1,  \ldots, S_K)^\top$ is a $K$-dimensional vector of sensitive attributes, such as gender or income,  $A \in \mathcal{A}:=\{1,-1\}$ denotes treatment and $R$ is  reward.  We assume $R$ is bounded and higher values of $R$ are desired. An ITR or policy $\mathcal{D}(\boldsymbol{X},\boldsymbol{S})$ is a function from feature space $\mathcal{X}\times\mathcal{S}$ to $\mathcal{A}$. For simplicity, let $(\boldsymbol{X}, \boldsymbol{S})$ denote a $p+K$ dimensional column vector $(\boldsymbol{X}^\top, \boldsymbol{S}^\top)^\top$.

The optimal ITR can be expressed as a rule that maximizes the value function $\mathcal{V}(\mathcal{D})$ \citep{zhao2012estimating}, defined as
\begin{equation}
\label{e1}
\begin{aligned}
\mathcal{V}(\mathcal{D})=E^{\mathcal{D}}(R)=E\left[\frac{I(A=\mathcal{D}(\boldsymbol{X},\boldsymbol{S}))}{A \pi+(1-A) / 2} R\right],
\end{aligned}
\end{equation}
where $I(\cdot)$ is the indicator function and $\pi=P(A=1)$.
Then, an optimal ITR, $\mathcal{D}^*$ is a rule that maximizes $\mathcal{V}(\mathcal{D})$, i.e.,
$
\mathcal{D}^* \in \underset{\mathcal{D}}{\arg \max }
   \mathcal{V}(\mathcal{D}(\boldsymbol{X}, \boldsymbol{S})).
$
It is worth noting that $\boldsymbol{S}$ can influence $\mathcal{D}^*$, resulting in different treatments being assigned to individuals who share identical characteristics but belong to different sensitive attribute groups. Without loss of generality, we assume that $R$ is nonnegative.


\subsection{A motivating example}
  To illustrate the concept of unfairness, we present a toy example.
In a non-profit entrepreneurship training program, some applicants are admitted  to participate for free. Our objective  is to maximize the innovation score $R$ that each applicant receives from a panel. Assume that the subjective score $R$ is influenced by both the individual's characteristic  $X$  and their gender $S$, modeled as:
\begin{equation}
\label{e2}
\begin{aligned}
R=1+X+0.5(5 X-3 S-3 )(A+1)+\varepsilon,
\end{aligned}
\end{equation}
where covariate  $X \sim$ Uniform[0,1], sensitive attribute $S \sim$ Bernoulli$(0.5)$, noise $\epsilon \sim$ Normal$(0,1)$. $A=1$ if the applicants is admitted and $A=-1$ otherwise. Moreover, we assume that $S=1$ for female and $S=0$ for male.

From Equation (\ref{e2}), it becomes evident that for  admitted individuals, the average reward for females $E(R|X, S=1, A=1)=6X-5$ is lower than that for males $E(R|X, S=0, A=1)=6X-2$.
Additionally, the optimal ITR can be calculated as follows:
 $$\mathcal{D}^*(X,{S})=
\begin{cases}
2I(X>\frac{3}{5})-1, & \text{if }  S = 0, \\
-1, & \text{if } S = 1. \\
\end{cases}$$

The optimal ITR $\mathcal{D}^*(X, S)$ shown above is an unfair policy because it treats individuals with the same covariate value differently based on gender.  Specifically, $\mathcal{D}^*(X={x},S=1)\neq \mathcal{D}^*(X={x},S=0)$ when $1 \geq x >3 / 5$ solely due to different gender identities. Consequently, according to the optimal ITR $\mathcal{D}^*(X, S)$, all females will not be admitted while   $60\%$ of the males will  be admitted. Therefore, this policy is gender biased and can raise serious legal and ethical concerns if implemented in the  welfare system.

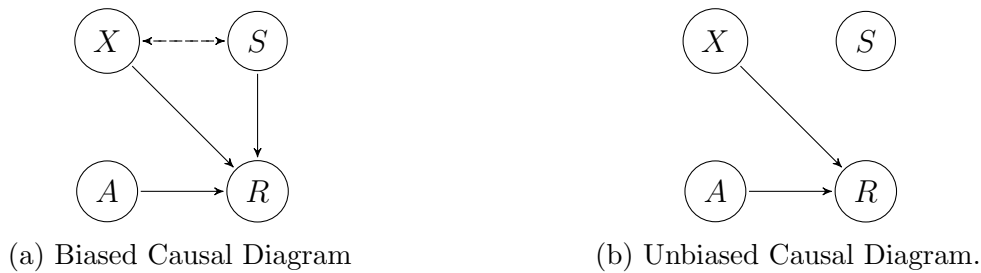
\begin{figure}[h]
    \centering
    \begin{minipage}{0.45\linewidth}
        \centering
        \begin{tikzpicture}[node distance=2cm, auto]
            \tikzset{
                block/.style = {draw,circle, minimum width=0.5cm, minimum height=0.5cm},
                arrow/.style = {->, >=stealth', shorten >=1pt, shorten <=1pt}
            }
            \node[block] (A) {$A$};
            \node[block, above of=A] (X) {$X$};
            \node[block, right of=A] (R) {$R$};
            \node[block, above of=R] (S) {$S$};
            \draw[arrow] (A) -- (R);
            \draw[arrow] (X) -- (R);
            \draw[arrow] (S) -- (R);
            \draw[dashed, arrow] (X) -- (S);
            \draw[dashed, arrow] (S) -- (X);
        \end{tikzpicture}
       \subcaption{Biased Causal Diagram}
    \label{fig2}
    \end{minipage}
    \hspace{0.5cm}
    \begin{minipage}{0.45\linewidth}
        \centering
        \begin{tikzpicture}[node distance=2cm, auto]
            \tikzset{
                block/.style = {draw,circle, minimum width=0.5cm, minimum height=0.5cm},
                arrow/.style = {->, >=stealth', shorten >=1pt, shorten <=1pt}
            }
            \node[block] (A) {$A$};
            \node[block, above of=A] (X) {$X$};
            \node[block, right of=A] (Y) {$R$};
            \node[block, above of=Y] (S) {$S$};
            \draw[arrow] (A) -- (Y);
            \draw[arrow] (X) -- (Y);
        \end{tikzpicture}
        \subcaption{Unbiased Causal Diagram.}
        \label{fig3}
    \end{minipage}
    \caption{Comparison of biased causal diagram and unbiased causal diagram.}
\end{figure}

Figure \ref{fig2} presents a causal diagram that exhibits bias, indicating two potential sources of unfairness in the policy. Firstly, with biased data, the sensitive attribute $S$ can influence the reward $R$. Secondly, the correlation between the covariate $X$ and the sensitive attribute $S$ can also impact the reward $R$, resulting in an unfair policy. Conversely, Figure \ref{fig3} illustrates an ideal unbiased causal diagram, where the covariate and the reward are independent of the sensitive attribute.

\subsection{Demographic parity individualized treatment rules}
 In our setting, we define fairness as a policy that ensures equal treatment for all individuals, regardless of their sensitive attributes. Therefore, we define the    Demographic Parity ITR as follows.
\begin{definition}
(Demographic Parity ITR). A policy $\mathcal{D}: \mathcal{X}\times \mathcal{S}\rightarrow \mathcal{A}$ attains    demographic parity under $(\boldsymbol{X}, \boldsymbol{S}, A, R)$, if  $\mathcal{D}(\boldsymbol{X},\boldsymbol{S})$ is independent of the sensitive attributes $\boldsymbol{S}$, i.e., $\mathcal{D}(\boldsymbol{X}, \boldsymbol{S})\perp \boldsymbol{S}$.
\end{definition}
A   Demographic Parity ITR  is equivalent to
\begin{equation}
\label{e3}
\begin{aligned}
{P}[\mathcal{D}(\boldsymbol{X},\boldsymbol{S})=a \mid \boldsymbol{S}=\boldsymbol{s}]={P}[\mathcal{D}(\boldsymbol{X},\boldsymbol{S})=a\mid \boldsymbol{S}=\boldsymbol{s}^{\prime}],
\end{aligned}
\end{equation}
for any $ \boldsymbol{s},\boldsymbol{s}^{\prime} \in \mathcal{S}$ and $a \in \{-1,1\}$.
Set the policy $\mathcal{D}(\boldsymbol{X},\boldsymbol{S})=\text{Sgn}[f(\boldsymbol{X},\boldsymbol{S})]$, where $f(\cdot)$ is a decision function
and $\text{Sgn}(u): =2I(u>0)-1$. It follows that Equation (\ref{e3}) is equivalent to
\begin{equation}
\label{e4}
\begin{aligned}
{P}[f(\boldsymbol{X},\boldsymbol{S})>0 \mid \boldsymbol{S}=\boldsymbol{s}]={P}[f(\boldsymbol{X},\boldsymbol{S})>0\mid \boldsymbol{S}=\boldsymbol{s}^{\prime}].
\end{aligned}
\end{equation}
Following Equation (\ref{e4}), we formulate the    Demographic Parity ITR by maximizing the value function while incorporating a demographic parity constraint as follows:
\begin{equation}
\label{e5}
\begin{aligned}
\underset{f}{\text{maximize }} \mathcal{V}(\mathcal{D}(f)),
 \text{ subject to } {P}[f(\boldsymbol{X}, \boldsymbol{S})>0 \mid \boldsymbol{S}=\boldsymbol{s}]={P}[f(\boldsymbol{X}, \boldsymbol{S})>0\mid \boldsymbol{S}=\boldsymbol{s}^{\prime}]
\end{aligned}
\end{equation}
for any $\boldsymbol{s},\boldsymbol{s}' \in \mathcal{S}$, where the constraint can be directly estimated through the empirical distribution of $f(\boldsymbol{X}, \boldsymbol{S})$ conditional on $ \boldsymbol{S}$.

\section{Methodology}

To find    Demographic Parity ITR, one straightforward approach is to ensure that $f(\boldsymbol{X}, \boldsymbol{S}) \perp \boldsymbol{S}$, satisfying Equation (\ref{e4}). However, solving Equation (\ref{e5}) becomes exceedingly challenging due to the non-convex and discontinuous nature of the indicator function in the empirical distribution.

Thus, we propose utilizing two fairness proxies, namely a linear and a nonlinear proxy, as substitutes for the demographic parity constraint in Equation (\ref{e4}). We refer to the ITR that conforms to the fairness proxy as the Demographic Parity-aware ITR (DPA-ITR). Moreover, we will estimate DPA-ITR under the linear or nonlinear proxy constraint.

\subsection{Linear  fairness  proxy}

A natural approach to represent independence is to consider the covariance between the sensitive attributes $\boldsymbol{S}$ and the decision function $f(\boldsymbol{X}, \boldsymbol{S}).$ In this regard, we propose using covariance as the linear fairness constraint: 
\begin{equation}
\begin{aligned}
\operatorname{Cov}\left(\boldsymbol{S}, f(\boldsymbol{X}, \boldsymbol{S})\right)
=E\left[(\boldsymbol{S}-E\boldsymbol{S}) f(\boldsymbol{X}, \boldsymbol{S})\right].
\nonumber
\end{aligned}
\end{equation}

Consider an observed data set $\left\{\left(\boldsymbol{X}_i, \boldsymbol{S}_i, A_i, R_i\right)\right\}_{i=1}^n$, where $\boldsymbol{S}_i=(S_{i1},  \ldots, S_{iK})$ and $S_{ik}$ represents the $k$-th sensitive attribute for the $i$th individual. Let $\overline{S}_{\cdot k}=\sum_{i=1}^n{S}_{ik}/n$ denote the sample average of the $k$th sensitive attribute across the sample.
To estimate the linear fairness proxy $\operatorname{Cov}\left(\boldsymbol{S}, f(\boldsymbol{X}, \boldsymbol{S})\right)$, we employ the following estimator:
\begin{equation}
\label{e6}
\begin{aligned}
\operatorname{\widehat{Cov}}\left(\boldsymbol{S}, f(\boldsymbol{X}, \boldsymbol{S})\right)
= \frac{\sum_{i=1}^n(\boldsymbol{S}_{i}-\boldsymbol{\overline{S}})f(\boldsymbol{X}_i,\boldsymbol{S}_i)}{n},
\nonumber
\end{aligned}
\end{equation}
where $\boldsymbol{\overline{S}}=\sum_{i=1}^n\boldsymbol{S}_{i}/n$ is the sample mean of the sensitive attributes $\boldsymbol{S}$. 
Here, the $k$-th element of the estimator is
$$
\frac{\sum_{i=1}^n({S}_{ik}-\overline{S}_{\cdot k})f(\boldsymbol{X}_i,\boldsymbol{S}_i)}{n}.
$$

\subsection{Nonlinear fairness  proxy}
It is well-established that zero covariance between two variables does not imply their independence. Hence, linear fairness proxy equal to zero does not guarantee fairness in all cases. To illustrate this with an example, we consider the following joint distribution of $S$ and $f(\boldsymbol{X}, {S})$:

\vspace{0.2in}
  \begin{center}
\begin{tabular}{cccc}
  \toprule
    Joint probability & $S=-1 $ & $S=0$& $S=1$\\
      \midrule
     $ f(\boldsymbol{X}, {S})=1$ & 0.25  & 0 &0.25  \\
  \midrule
   $ f(\boldsymbol{X}, {S})=-1$  & 0 &0.5  & 0 \\
  \bottomrule
\end{tabular}
\end{center}
\vspace{0.2in}

Evidently, the decision function varies with different values of sensitive attribute $S$, indicating an unfair policy. However, the linear fairness proxy $\operatorname{Cov}\left({S}, f(\boldsymbol{X}, {S})\right)$ equals to zero, leading to contradictions. For this reason, we propose a more general nonlinear fairness proxy that can capture the nonlinear correlations. Inspired by \cite{zhuliping2011}, we present a nonlinear covariance as follows
\begin{equation}
\begin{aligned}
\boldsymbol{\Omega}(\boldsymbol{s},f)=E[ f(\boldsymbol{X}, \boldsymbol{S})\left\{\boldsymbol{I}(\boldsymbol{S}<\boldsymbol{s})-E\boldsymbol{I}(\boldsymbol{S}<\boldsymbol{s})\right\}],
\nonumber
\end{aligned}
\end{equation}
where $\boldsymbol{s}=(s_1,  \ldots, s_K)$, $I(\cdot)$ is the indicator function and $\boldsymbol{I}(\boldsymbol{S}<\boldsymbol{s}) $ represents $(I({S}_1<{s}_1), I({S}_2<{s}_2), \cdots, I({S}_K<{s}_K) )^\top$. The set of indicator functions $\{ {I}({S}<s) \mid s \in \mathcal{S} \}$ offers a more comprehensive approach to capturing nonlinear correlations compared to the limitations of linear covariance.

Let $\Omega_k({s}_k,f)$ denote the $k$-th element of $\boldsymbol{\Omega}(\boldsymbol{s},f)$. Define
$$
\omega_k(f)=E\left\{\Omega_k({S}_k,f)\right\}, \quad k=1, \ldots, K,
$$
where $\omega_k(f)$ is the correlation measure. If $f(\boldsymbol{X}, \boldsymbol{S})$ and ${S}_k$ are independent, it follows that $\Omega_k({s},f)=0$ for any ${s} \in \mathcal{R}$, thereby resulting in $\omega_k(f)=0$. Conversely, if $f(\boldsymbol{X}, \boldsymbol{S})$ and ${S}_k$ are correlated, then there exists some ${s}$ for which $\Omega_k(s,f) \neq 0$. Hence, $\boldsymbol{\omega}(f)=(\omega_1(f),  \ldots, {\omega}_K(f))^\top$ is utilized as a nonlinear fairness proxy. In the given example, the nonlinear fairness proxy $\omega_1(f)$ is calculated as $-\frac{1}{8}$ using the joint distribution, effectively capturing the correlation.


Let $\widehat{\operatorname{\boldsymbol{\Omega}}}(\boldsymbol{s},f)$ be the estimator of $\boldsymbol{\Omega}(\boldsymbol{s},f)$, defined by
$$
\widehat{\operatorname{\boldsymbol{\Omega}}}(\boldsymbol{s},f)
= \frac{\sum_{i=1}^n(\boldsymbol{I}(\boldsymbol{S}_i<\boldsymbol{s})-\overline{\boldsymbol{I}}(\boldsymbol{S}<\boldsymbol{s}))f(\boldsymbol{X}_i,\boldsymbol{S}_i)}{n},
$$
where $\boldsymbol{I}(\boldsymbol{S}_i<\boldsymbol{s})=(I({S}_{i1}<{s}_1), I({S}_{i2}<{s}_2), \cdots, I({S}_{iK}<{s}_K) )^\top$ and $\overline{\boldsymbol{I}}(\boldsymbol{S}<\boldsymbol{s})=\frac{1}{n}\sum_{i=1}^n\boldsymbol{I}(\boldsymbol{S}_i<\boldsymbol{s})$.
Let $\widehat{\Omega}_k(s_k,f)$ be the $k$-th element of $\widehat{\operatorname{\boldsymbol{\Omega}}}(s,f)$, where
$$\widehat{\Omega}_k(s_k,f)=\frac{\sum_{i=1}^n(I({S}_{ik}<{s_k})-\overline{{I}}({S}_{\cdot k}<{s_k}))f(\boldsymbol{X}_i,\boldsymbol{S}_i)}{n}$$
and $\overline{{I}}({S}_{\cdot k}<{s}_k)=\frac{1}{n}\sum_{i=1}^nI({S}_{ik}<{s}_{k})$. Let $\boldsymbol{\widehat{\omega}}(f)$ be the estimator of $\boldsymbol{\omega}(f)$, defined by $\boldsymbol{\widehat{\omega}}(f)=(\widehat{\omega}_1(f),  \ldots, \widehat{\omega}_K(f))^\top$ with $\widehat{\omega}_k(f)=\frac{\sum_{j=1}^n\widehat{{\Omega}}_k(S_{jk},f)}{n}$.

Next, we provide solutions of optimal DPA-ITR for both linear and nonlinear fairness proxies.

\subsection{Estimation of DPA-ITR under the linear fairness proxy}
Since solving the optimization problem in Expression (\ref{e5}) can be challenging, we utilize the following fairness proxy,
\begin{equation}
\label{e7}
\begin{aligned}
\underset{f}{ \text{maximize }}\mathcal{V}(\mathcal{D}(f)),
 \text { subject to } |{\operatorname{Cov}\left(\boldsymbol{S}, f(\boldsymbol{X}, \boldsymbol{S})\right)}| \leq \boldsymbol{c},
\end{aligned}
\end{equation}
where $\boldsymbol{c}=(c_1,\ldots,c_K)^\top$ denotes the fairness constraint and can be selected through empirical experiments.

Motivated by \cite{zhao2012estimating}, maximizing $\mathcal{V}(\mathcal{D}(f))$ is equivalent to minimizing the following function:
\begin{equation}
\label{e8}
\begin{aligned}
E[R \mid A=1]+E[R \mid A=-1] -\mathcal{V}(\mathcal{D}(f))
 =E\left[\frac{I(A \neq \mathcal{D}(f(\boldsymbol{X}, \boldsymbol{S})))}{A \pi+(1-A) / 2} R\right].
\end{aligned}
\end{equation}
Given the observed data $\left\{\left(\boldsymbol{X}_i, \boldsymbol{S}_i, A_i, R_i\right)\right\}_{i=1}^n$, we have the empirical version of (\ref{e7}) with Equation (\ref{e8}) as follows:
\begin{equation}
\label{e9}
\begin{aligned}
\underset{f}{\operatorname{minimize}}\quad &n^{-1} \sum_{i=1}^n \frac{R_i}{A_i \pi+\left(1-A_i\right) / 2} I\left(A_i \neq \operatorname{\text{Sgn}}\left(f\left(\boldsymbol{X}_i,\boldsymbol{S}_i\right)\right)\right), \\
 \text {subject to } &|{\operatorname{\widehat{Cov}}\left(\boldsymbol{S}, f(\boldsymbol{X}, \boldsymbol{S})\right)}| \leq \boldsymbol{c}.
\end{aligned}
\end{equation}
To address the optimization challenges posed by the non-smooth indicator function in Equation (\ref{e9}), we employ a convex hinge surrogate \citep{zhao2012estimating}. This surrogate incorporates a ridge penalty, resulting in a regularized version of the optimization problem:
\begin{equation}
\label{e10}
\begin{aligned}
\underset{f}{\operatorname{minimize}}\quad &n^{-1} \sum_{i=1}^n \frac{R_i}{A_i \pi+\left(1-A_i\right) / 2}\left(1-A_i f\left(\boldsymbol{X}_i,\boldsymbol{S}_i\right)\right)^{+}+\lambda\|f\|^2 \text {, } \\
 \text {subject to } &-{\operatorname{\widehat{Cov}}\left(\boldsymbol{S}, f(\boldsymbol{X}, \boldsymbol{S})\right)} \geq \boldsymbol{c},  \text
 { and }
{\operatorname{\widehat{Cov}}\left(\boldsymbol{S}, f(\boldsymbol{X}, \boldsymbol{S})\right)} \leq \boldsymbol{c},
\end{aligned}
\end{equation}
where $u^{+}=\max (u, 0) $ and $\|f\| $ is some norm for $ f $.

Next, we consider linear and nonlinear forms for decision function $f$.

\subsection{Linear decision function for DPA-ITR estimation}
Suppose that the decision function $f(\cdot)$ takes a linear form, such that $f(\boldsymbol{X},\boldsymbol{S})=\langle\boldsymbol{\beta}, (\boldsymbol{X},\boldsymbol{S})\rangle+$ ${\beta}_0$, where $\langle\cdot, \cdot\rangle$ denotes the inner product in Euclidean space and $\|f\|=\sqrt{\langle\boldsymbol{\beta}, \boldsymbol{\beta}\rangle}$. Then the ITR will assign treatment value 1 if $\langle\boldsymbol{\beta}, (\boldsymbol{X},\boldsymbol{S})\rangle+{\beta}_0>0$ and -1 otherwise. Moreover, we have $\left(1-A_i f\left(\boldsymbol{X}_i,\boldsymbol{S}_i\right)\right)^{+}=\left(1- A_i[\left\langle\boldsymbol{\beta},(\boldsymbol{X}_i, \boldsymbol{S}_i) \right\rangle+{\beta}_0]\right)^{+}$.

Let $\xi_i=\left[1- A_i\{\left\langle\boldsymbol{\beta},(\boldsymbol{X}_i, \boldsymbol{S}_i) \right\rangle+{\beta}_0\}\right]^{+}$. Then the optimization problem (\ref{e10}) becomes (see details in Supplementary Materials E):
\begin{equation}
\label{e11}
\begin{aligned}
\underset{\boldsymbol{\beta}, {\beta}_0,\xi}{\operatorname{minimize}}\quad&\frac{1}{2}\|\boldsymbol{\beta}\|^2+\kappa \sum_{i=1}^n \frac{R_i}{\pi_i} \xi_i \quad, \\
\text { subject to } & A_i[\left\langle\boldsymbol{\beta},(\boldsymbol{X}_i, \boldsymbol{S}_i) \right\rangle+{\beta}_0]\geq \left(1-\xi_i\right),\\
&\xi_i \geq 0,\\
-&\frac{\sum_{i=1}^n(\boldsymbol{S}_{i}-\overline{\boldsymbol{S}})[\left\langle\boldsymbol{\beta}, (\boldsymbol{X}_i, \boldsymbol{S}_i) \right\rangle+{\beta}_0]}{n} \leq \boldsymbol{c},\\
&\frac{\sum_{i=1}^n(\boldsymbol{S}_{i}-\overline{\boldsymbol{S}})[\left\langle\boldsymbol{\beta},(\boldsymbol{X}_i, \boldsymbol{S}_i) \right\rangle+{\beta}_0]}{n} \leq  \boldsymbol{c},
\end{aligned}
\end{equation}
where $i=1,\ldots, n$ and  $\pi_i=\pi I\left(A_i=1\right)+(1-\pi) I\left(A_i=-1\right)$ and $\kappa=\frac{1}{2\lambda}>0$ is a tuning parameter.
To transform the optimization problem in Expression (\ref{e11}), we propose the Lagrange function:
\begin{equation}
\label{e12}
\begin{aligned}
& \frac{1}{2}\|\boldsymbol{\beta}\|^2+\kappa \sum_{i=1}^n \frac{R_i}{\pi_i} \xi_i \\
-&\sum_{i=1}^n \alpha_i\left\{A_i\left((\boldsymbol{X}_i,\boldsymbol{S}_i)^\top \boldsymbol{\beta}+\beta_0\right)-\left(1-\xi_i\right)\right\}-\sum_{i=1}^n \mu_i \xi_i\\
+&\sum_{k=1}^K\frac{\gamma_k }{n}\sum_{i=1}^n\left\{-({S}_{ik}-\overline{S}_{\cdot k})\left((\boldsymbol{X}_i,\boldsymbol{S}_i)^\top \boldsymbol{\beta}+\beta_0\right) - c_k \right\}\\
+&\sum_{k=1}^K\frac{\eta_k }{n}\sum_{i=1}^n\left\{({S}_{ik}-\overline{S}_{\cdot k})\left((\boldsymbol{X}_i,\boldsymbol{S}_i)^\top \boldsymbol{\beta}+\beta_0\right) - c_k \right\}
\nonumber
\end{aligned}
\end{equation}
with $\alpha_i \geq 0, \mu_i \geq 0, \gamma_k\geq 0,\eta_k\geq 0,  i=1,\ldots, n, k=1,\ldots,K$. Taking derivatives with respect to $\left(\boldsymbol{\beta}, \beta_0\right)$ and $\xi_i$, we have
\begin{equation}
\label{e13}
\begin{aligned}
\boldsymbol{\beta}&=\sum_{i=1}^n \alpha_i A_i (\boldsymbol{X}_i,\boldsymbol{S}_i)+\sum_{k=1}^K\frac{\gamma_k }{n}\sum_{i=1}^n({S}_{ik}-\overline{S}_{\cdot k})(\boldsymbol{X}_i,\boldsymbol{S}_i)\\
&-\sum_{k=1}^K\frac{\eta_k }{n}\sum_{i=1}^n({S}_{ik}-\overline{S}_{\cdot k})(\boldsymbol{X}_i,\boldsymbol{S}_i) \\
&=\sum_{i=1}^n[\alpha_i A_i +\sum_{k=1}^K\frac{(\gamma_k-\eta_k)}{n}({S}_{ik}-\overline{S}_{\cdot k})] (\boldsymbol{X}_i,\boldsymbol{S}_i) \\
0&=\sum_{i=1}^n \alpha_i A_i+\sum_{k=1}^K\frac{\gamma_k }{n}\sum_{i=1}^n({S}_{ik}-\overline{S}_{\cdot k})- \sum_{k=1}^K\frac{\eta_k }{n}\sum_{i=1}^n({S}_{ik}-\overline{S}_{\cdot k})\\
\alpha_i&=\kappa R_i / \pi_i-\mu_i, i=1,\ldots, n.
\nonumber
\end{aligned}
\end{equation}
Plugging these equations into the Lagrange function, we obtain
\begin{equation}
\begin{aligned}
-\frac{1}{2}\|\boldsymbol{\beta}\|^2
+\sum_{i=1}^n \alpha_i
-\sum_{k=1}^Kc_k\gamma_k
-\sum_{k=1}^Kc_k\eta_k.
\nonumber
\end{aligned}
\end{equation}
Then, the dual problem is
\begin{equation}
\label{14}
\begin{aligned}
&\underset{\boldsymbol{\alpha},  \boldsymbol{\gamma},  \boldsymbol{\eta}}{\operatorname{maximize}}
\sum_{i=1}^n \alpha_i
-\sum_{k=1}^Kc_k\gamma_k
-\sum_{k=1}^Kc_k\eta_k\\
&-\frac{1}{2}\sum_{i=1}^n\sum_{j=1}^n[\alpha_i A_i +\sum_{k=1}^K\frac{(\gamma_k-\eta_k)}{n}({S}_{ik}-\overline{S}_{\cdot k})]\\
&*[\alpha_j A_j +\sum_{k=1}^K\frac{(\gamma_k-\eta_k)}{n}(S_{jk}-\overline{S}_{\cdot k})]\\
&* \langle(\boldsymbol{X}_i,\boldsymbol{S}_i),(\boldsymbol{X}_j,\boldsymbol{S}_j) \rangle
\end{aligned}
\end{equation}
subject to $0\leq \gamma_k, 0\leq\eta_k$ for $ k=1,\ldots, K$ and $0\leq \alpha_i \leq \kappa R_i / \pi_i$ for $ i=1, \ldots, n$.
This dual problem can be  solved via quadratic programming software packages.
Finally, we obtain that $ \widehat{\boldsymbol{\beta}}=\sum_{i=1}^n[\widehat{\alpha}_i A_i +\sum_{k=1}^K\frac{(\widehat{\gamma}_k-\widehat{\eta}_k)}{n}({S}_{ik}-\overline{S}_{\cdot k})] (\boldsymbol{X}_i,\boldsymbol{S}_i)$ and $\widehat{\beta}_0$ can be solved using the margin point $0<\widehat{\alpha}_m, \widehat{\xi}_m=0$ subject to the Karush-Kuhn-Tucker conditions.

\subsection{Nonlinear decision function for DPA-ITR estimation}

To overcome the challenges posed by high dimensionality and linear inseparability \citep{zhao2012estimating}, we incorporate kernel functions to introduce nonlinearity into the decision function. Taking a kernel function $\mathcal{K}\langle \cdot,\cdot \rangle$ that satisfies continuity, symmetry, and positive semidefiniteness, we construct the reproducing kernel Hilbert space (RKHS) $\mathcal{H}_{\mathcal{K}}$ as the completion of the linear span of all functions $\{\mathcal{K}\langle \cdot,(\boldsymbol{X},\boldsymbol{S)}\rangle| (\boldsymbol{X},\boldsymbol{S)}) \in \mathcal{X}\times\mathcal{S}\}$ (see Moore–Aronszajn Theorem in \cite{aronszajn1950theory}). Let $\|\cdot\|_{\mathcal{K}}$ denote the norm in $\mathcal{H}_{\mathcal{K}}$, which can be induced by the following inner product,
$$
\langle f, g\rangle=\sum_{i=1}^n \sum_{j=1}^n h_i \beta_j \mathcal{K}\langle(\boldsymbol{X}_i,\boldsymbol{S}_i),(\boldsymbol{X}_j,\boldsymbol{S}_j)\rangle,
$$
for $f(\cdot)=\sum_{i=1}^n h_i \mathcal{K}\langle \cdot, (\boldsymbol{X}_i,\boldsymbol{S}_i)\rangle$ and $g(\cdot)=\sum_{j=1}^n \beta_j \mathcal{K}\langle \cdot, (\boldsymbol{X}_j,\boldsymbol{S}_j)\rangle$.
It follows that the optimal decision function is given by
$$
\sum_{i=1}^n[\widehat{\alpha}_i A_i +\sum_{k=1}^K\frac{(\widehat{\gamma}_k-\widehat{\eta}_k)}{n}({S}_{ik}-\overline{S}_{\cdot k})]  \mathcal{K}\langle \cdot, (\boldsymbol{X}_i,\boldsymbol{S}_i)\rangle+\widehat{b}_0,
$$
where $\left(\widehat{\alpha}_1, \ldots, \widehat{\alpha}_N,\widehat{\gamma}_1,\widehat{\eta}_1,  \ldots, \widehat{\gamma}_k,\widehat{\eta}_k \right)^\top$ can be obtained by  solving
\begin{equation}
\label{e15}
\begin{aligned}
&\underset{\boldsymbol{\alpha},  \boldsymbol{\gamma},  \boldsymbol{\eta}}{\operatorname{maximize}}
\sum_{i=1}^n \alpha_i
-\sum_{k=1}^Kc_k\gamma_k
-\sum_{k=1}^Kc_k\eta_k\\
&-\frac{1}{2}\sum_{i=1}^n\sum_{j=1}^n[\alpha_i A_i +\sum_{k=1}^K\frac{(\gamma_k-\eta_k)}{n}({S}_{ik}-\overline{S}_{\cdot k})]\\
&*[\alpha_j A_j +\sum_{k=1}^K\frac{(\gamma_k-\eta_k)}{n}(S_{jk}-\overline{S}_{\cdot k})]\\
&* \mathcal{K}\langle(\boldsymbol{X}_i,\boldsymbol{S}_i),(\boldsymbol{X}_j,\boldsymbol{S}_j) \rangle
\end{aligned}
\end{equation}
subject to $0\leq \gamma_k, 0\leq\eta_k, 0\leq \alpha_i \leq \kappa R_i / \pi_i, i=1, \ldots, n$, and $\widehat{b}_0$ can be solved using Karush-Kuhn-Tucker conditions. Specifically, for the $m$-th data point, if $0<\widehat{\alpha}_m$, then $\widehat{\xi}_m=0$, which means
{$$\widehat{b}_0= A_m-\sum_{i=1}^n\left[\widehat{\alpha}_i A_i +\sum_{k=1}^K\frac{(\widehat{\gamma}_k-\widehat{\eta}_k)}{n}\left({S}_{ik}-\overline{S}_{\cdot k}\right)\right]  \mathcal{K}\langle (\boldsymbol{X}_m, \boldsymbol{S}_m), (\boldsymbol{X}_i,\boldsymbol{S}_i)\rangle.$$}
If we choose $\mathcal{K}\langle z_1,z_2 \rangle =\langle z_1,z_2 \rangle$, then (\ref{e15}) reduces to (\ref{14}).

\subsection{Estimation of DPA-ITR under the nonlinear fairness proxy}

We employ the nonlinear fairness proxy and reformulate Expression (\ref{e5}) into the following optimization problem:
\begin{equation}
\begin{aligned}
\underset{f}{ \text{maximize }}\mathcal{V}(\mathcal{D}(f)),
 \text { subject to } |\boldsymbol{{\omega}}(f)| \leq \boldsymbol{c},
 \nonumber
\end{aligned}
\end{equation}
where $\boldsymbol{c}=(c_1,\ldots,c_K)^\top$ is the fairness constraint.

\subsection{Linear decision function for DPA-ITR estimation}
Similar to Section 3.2.1, for a linear decision function $f(\boldsymbol{x},\boldsymbol{s})=\langle\boldsymbol{\beta}, (\boldsymbol{x},\boldsymbol{s})\rangle+$ ${\beta}_0$, we introduce a convex hinge surrogate with a ridge penalty and subsequently derive its corresponding dual problem as follows:
\begin{equation}
\label{e16}
\begin{aligned}
&\underset{\boldsymbol{\alpha},  \boldsymbol{\gamma},  \boldsymbol{\eta}}{\operatorname{maximize}}
\sum_{i=1}^n \alpha_i
-\sum_{k=1}^Kc_k\gamma_k
-\sum_{k=1}^Kc_k\eta_k\\
&-\frac{1}{2}\sum_{i=1}^n\sum_{j=1}^n\left[\alpha_i A_i +\sum_{k=1}^K\frac{(\gamma_k-\eta_k)}{n}\left\{\frac{1}{n}\sum_{l=1}^n\left(I({S}_{ik}<S_{lk})-\overline{{I}}({S}_{\cdot k}<S_{lk})\right)\right\}\right]\\
&*\left[\alpha_j A_j +\sum_{k=1}^K\frac{(\gamma_k-\eta_k)}{n}\left\{\frac{1}{n}\sum_{l=1}^n\left(I({S}_{ik}<S_{lk})-\overline{{I}}({S}_{\cdot k}<S_{lk})\right)\right\}\right]\\
&* \langle(\boldsymbol{X}_i,\boldsymbol{S}_i),(\boldsymbol{X}_j,\boldsymbol{S}_j) \rangle,
\end{aligned}
\end{equation}
subject to $0\leq \gamma_k, 0\leq\eta_k$ for $ k=1,\ldots, K$ and $0\leq \alpha_i \leq \kappa R_i / \pi_i$ for $ i=1, \ldots, n$. Derivation details are provided in Supplementary Material A.

Finally, we obtain that
$$\qquad \widehat{\boldsymbol{\beta}}=\sum_{i=1}^n\left[\widehat{\alpha}_i A_i +\sum_{k=1}^K\frac{(\widehat{\gamma}_k-\widehat{\eta}_k)}{n}\left(\frac{1}{n}\sum_{l=1}^n\left(I\left({S}_{ik}<S_{lk}\right)-\overline{{I}}\left({S}_{\cdot k}<S_{lk}\right)\right)\right)\right] \left(\boldsymbol{X}_i,\boldsymbol{S}_i\right),$$
where $\widehat{\beta}_0$ can be solved using the margin point $0<\widehat{\alpha}_m, \widehat{\xi}_m=0$ subject to the Karush-Kuhn-Tucker conditions.

\subsection{Nonlinear decision function for DPA-ITR estimation}
For a nonlinear decision function $\widehat{f}(\cdot) \in \mathcal{H}_{\mathcal{K}}$, it can be shown that the optimal decision function is
$$
\sum_{i=1}^n\left[\widehat{\alpha}_i A_i +\sum_{k=1}^K\frac{(\widehat{\gamma}_k-\widehat{\eta}_k)}{n}\left(\frac{1}{n}\sum_{l=1}^n\left(I\left({S}_{ik}<S_{lk}\right)-\overline{{I}}\left({S}_{\cdot k}<S_{lk}\right)\right)\right)\right]  \mathcal{K}\langle \cdot, (\boldsymbol{X}_i,\boldsymbol{S}_i)\rangle+\widehat{b}_0,
$$
where $\left(\widehat{\alpha}_1, \ldots, \widehat{\alpha}_N,\widehat{\gamma}_1,\widehat{\eta}_1,  \ldots, \widehat{\gamma}_k,\widehat{\eta}_k \right)^\top$ can be obtained by  solving
\begin{equation}
\label{e17}
\begin{aligned}
&\underset{\boldsymbol{\alpha},  \boldsymbol{\gamma},  \boldsymbol{\eta}}{\operatorname{maximize}}
\sum_{i=1}^n \alpha_i
-\sum_{k=1}^Kc_k\gamma_k
-\sum_{k=1}^Kc_k\eta_k\\
&-\frac{1}{2}\sum_{i=1}^n\sum_{j=1}^n\left[\alpha_i A_i +\sum_{k=1}^K\frac{(\gamma_k-\eta_k)}{n}\left\{\frac{1}{n}\sum_{l=1}^n\left(I({S}_{ik}<S_{lk})-\overline{{I}}({S}_{\cdot k}<S_{lk})\right)\right\}\right]\\
&*\left[\alpha_j A_j +\sum_{k=1}^K\frac{(\gamma_k-\eta_k)}{n}\left\{\frac{1}{n}\sum_{l=1}^n\left(I({S}_{ik}<S_{lk})-\overline{{I}}({S}_{\cdot k}<S_{lk})\right)\right\}\right]\\
&*\mathcal{K} \langle(\boldsymbol{X}_i,\boldsymbol{S}_i),(\boldsymbol{X}_j,\boldsymbol{S}_j) \rangle
\end{aligned}
\end{equation}
subject to  $0\leq \gamma_k, 0\leq\eta_k, 0\leq \alpha_i \leq \kappa R_i / \pi_i, i=1, \ldots, n$, and $\widehat{b}_0$ can be solved using Karush-Kuhn-Tucker conditions. Specifically, for the $m$-th data point, if $0<\widehat{\alpha}_m$, then $\widehat{\xi}_m=0$, which yields {\small$$\widehat{b}_0= A_m-\sum_{i=1}^n\left[\widehat{\alpha}_i A_i +\sum_{k=1}^K\frac{(\widehat{\gamma}_k-\widehat{\eta}_k)}{n}\left\{\frac{1}{n}\sum_{l=1}^n\left(I\left({S}_{ik}<S_{lk}\right)-\overline{{I}}\left({S}_{\cdot k}<S_{lk}\right)\right)\right\}\right]  \mathcal{K}\langle (\boldsymbol{X}_m, \boldsymbol{S}_m), (\boldsymbol{X}_i,\boldsymbol{S}_i)\rangle.$$}

\section{Theoretical results}
Firstly, it is essential to establish the convexity of the proposed optimization problems.
\begin{theorem}
\label{t1}
\textbf{ \upshape (Algorithm Convexity)}
Rewrite
$\boldsymbol{\alpha}=(\alpha_{1},\ldots,\alpha_n,\gamma_1,\eta_1, \ldots, \gamma_K, \eta_K)^\top,$
$\boldsymbol{e}=(1,\ldots,1,-c_1,-c_1,  \ldots, -c_K, -c_K)^\top$.
The target functions of optimization problem in (\ref{e16}) and  (\ref{e17}) are respectively expressed as
\begin{equation}
\label{e18}
\begin{aligned}
&\underset{\boldsymbol{\alpha}}{\operatorname{maximize}} \quad\boldsymbol{\alpha}^\top\boldsymbol{e}-\frac{1}{2}\boldsymbol{\alpha}^\top\mathbb{D}\boldsymbol{\alpha},
\nonumber
\end{aligned}
\end{equation}
\begin{equation}
\label{e19}
\begin{aligned}
&\underset{\boldsymbol{\alpha}}{\operatorname{maximize}} \quad\boldsymbol{\alpha}^\top\boldsymbol{e}-\frac{1}{2}\boldsymbol{\alpha}^\top\mathbb{D}_K\boldsymbol{\alpha}.
\nonumber
\end{aligned}
\end{equation}
Then, the matrices $\mathbb{D}$ and $\mathbb{D}_k$ are positive semidefinite.
\end{theorem}
Theorem \ref{t1} imples that the optimization problems in (\ref{e16}) and  (\ref{e17}) can be solved using convex quadratic programming. Consequently, it is possible to obtain a practical and desirable solution. As illustrated in Table 1, the solution ensures that the estimated fairness proxy is lower than the fairness constraint $c$.

Let $\mathcal{R}_\phi(f)=E\left[\frac{R\phi(Af(\boldsymbol{X}, \boldsymbol{S}))}{A \pi+(1-A) / 2} \right]$  denote $\phi$-risk, where $\phi(x)=(1-x)^{+}$, and let $\mathcal{R}(f)=E\left[\frac{I(A \neq \mathcal{D}(f(\boldsymbol{X}, \boldsymbol{S})))}{A \pi+(1-A) / 2} R\right]$  denote risk.
Here,  $\widehat{f_n}$  is given by
\begin{equation}\label{e20}
\begin{aligned}
\widehat{f}_n=&\underset{f \in \mathcal{H}_k}{\operatorname{argmin}}\left\{\frac{1}{n} \sum_{i=1}^n \frac{R_i}{\pi_i}\left\{1-A_i f\left(\boldsymbol{X}_i,\boldsymbol{S}_i\right)\right\}^{+}+\lambda_n\|f\|_k^2\right\},\\
&\text{subject to }|{\operatorname{\widehat{Cov}}\left(\boldsymbol{S}, f(\boldsymbol{X}, \boldsymbol{S})\right)}| \leq \boldsymbol{c}+C / \sqrt{n},
\nonumber
\end{aligned}
\end{equation}
where $C$ is a positive constant. The introduction of the term $C / \sqrt{n}$ is aimed at ensuring the feasibility of ${\operatorname{\widehat{Cov}}\left(\boldsymbol{S}, f(\boldsymbol{X}, \boldsymbol{S})\right)}$ for the sake of simplicity in theoretical exposition, which can be dropped in practice.

Next, we establish the consistency of the estimator $\widehat{f}_n$ by demonstrating that it asymptotically achieves the optimal $\phi$-risk under the fairness proxy constraint. To establish this consistency, we utilize empirical process techniques and provide the detailed proof in Supplementary Materials C.
\begin{theorem}\textbf{\upshape (Consistency)}
Assume that we choose a sequence $\lambda_n>0$, such that $\lambda_n \rightarrow 0$ and $\lambda_n n \rightarrow \infty$. Then for all distributions $P$, we have
$$
\\ \lim _{n \rightarrow \infty}\left\{\mathcal{R}_\phi\left(\widehat{f_n}\right)-\inf _{f \in \overline{\mathcal{F}}_c} \mathcal{R}_\phi(f)\right\}=0 \; \text{ in probability},
$$
where $\overline{\mathcal{F}}_c$ denotes the closure of $\mathcal{F}_c=\{f: |\operatorname{Cov}\left(\boldsymbol{S}, f(\boldsymbol{X}, \boldsymbol{S})\right)|\leq \boldsymbol{c} , f\in \mathcal{H}_{\mathcal{K}} \}$. \end{theorem}
Regarding the constraint $\mathcal{W}c=\{f: |\boldsymbol{\omega}(f)|\leq \boldsymbol{c} , f\in \mathcal{H}_{\mathcal{K}} \}$, consistency can be achieved as well, since $\boldsymbol{\widehat{\omega}}(f)$ converges to $\boldsymbol{\omega}(f)$ in probability (as stated in Theorem 2 of \cite{zhuliping2011}). Furthermore, under the assumption of geometric noise given by \cite{steinwart2007fast}, we can derive the convergence rate.

Suppose that $\mathcal{R}_{\phi}^{*}= \inf  \{\mathcal{R}_{\phi}(f)| f: \mathcal{X}\times\mathcal{S}\rightarrow \mathbb{R} \}$ and there exits a $f^{*}\in \mathcal{H}_{\mathcal{K}}$ such that $\mathcal{R}_{\phi}^{*}=R_{\phi}(f^{*})$. Based on Proposition 3.1  \citep{zhao2012estimating}, we have $\mathcal{R}^{*}=R(f^{*})$, where $\mathcal{R}^{*}= \inf  \{\mathcal{R}(f)| f: \mathcal{X}\times\mathcal{S}\rightarrow \mathbb{R} \}$.
\begin{theorem} \textbf{\upshape (Risk Bound)} Suppose that a distribution $P$ of $(\boldsymbol{X}, \boldsymbol{S}, A, R)$ satisfies condition ``geometric noise'' assumption (Supplementary Materials F) with noise exponent $q>0$ \citep{steinwart2007fast,zhao2012estimating} and  the bandwidth of Gaussian kernel $\sigma=\lambda_n^{-1 /(q+1)(p+K)} $. Then for any $\tau>1, \delta>0,0<v\leq2,$ $
\widehat{f}_n\in\mathcal{F}_c$,  we have
$$
P\left(\mathcal{R}(\widehat{f}_n)-\mathcal{R}^{*} \leq
C\epsilon +\inf _{f \in \mathcal{F}_{c}}\left[\mathcal{R}_\phi(f)-\mathcal{R}_\phi^{*}\right]\right)\geq 1-e^{-\tau}.
$$
where $\epsilon=\left(\lambda_n\right)^{-\frac{2}{2+\nu}+\frac{(2-\nu)(1+\delta)}{(2+\nu)(1+q)}} n^{-\frac{2}{2+\nu}}+\frac{\tau}{n \lambda_n}+B\lambda_n$ and $C$ is a constant that depends on $v,\delta, K$ and $\pi$.
\end{theorem}
 We can set $\lambda_n=n^{-\frac{2(1+q)}{(4+v) q+2+(2-v)(1+\delta)}}$ (see Theorem 2.7 in \cite{steinwart2007fast} and Theorem 3.4 in \cite{zhao2012estimating},) and $v\leq 1$.  Then, the risk can be expressed as
$\mathcal{R}(\widehat{f}_n)-\mathcal{R}^{*}=O_{p}(n^{-\frac{2(1+q)}{(4+v) q+2+(2-v)(1+\delta)}})+\inf _{f \in \mathcal{F}_{c}}\left[\mathcal{R}_\phi(f)-\mathcal{R}_\phi^{*}\right]$.
When the data is distinctly separable, it suggests that the parameter $q$ is sufficiently large. Moreover, assuming that both $\delta$ and $\nu$ are significantly small, the term $\epsilon$ becomes close to $O_p(n^{-\frac{1}{2}})$. As a result, the risk is primarily influenced by the fairness constraint.

\section{Simulation studies}

\subsection{Simulation designs}

We conduct extensive simulations to demonstrate the performance of the proposed DPA-ITR (fairness policy). In simulations, the treatment variable $A$ is independently generated from $\{-1,1\}$ with equal probabilities for each value. The response variable $R$ is assumed to follow a normal distribution with a mean of $T(\boldsymbol{X}, \boldsymbol{S},A)$ and a variance of $1$, where $T(\boldsymbol{X}, \boldsymbol{S},A)$ represents the interaction between the treatment and prognostic variables. For all four experiments, the covariate vector $\boldsymbol{X}$ is independently generated from Uniform$(-5,5)$, with dimensions $p=3,50$. For performance comparisons, we considered the OWL policy introduced by \cite{zhao2012estimating}.

\noindent\textbf{Experiment\ 1:}
We consider mean of the reward taking the form:
\begin{equation*}
\begin{aligned}
T(\boldsymbol{X}, S,A)=10+ X_1+X_2+0.25 X_3+\left(X_1+X_2-10SI(A=1)\right)A.
\end{aligned}
\end{equation*}


Assume that the sensitive attribute $S$ follows a Bernoulli distribution with parameter $p=\rho(X_1+X_2)$, where $\rho(u)=e^{u}/(1+e^u)$ is an increasing function. This implies that there exists a positive correlation between $S$ and the sum of $X_1$ and $X_2$.

\noindent\textbf{Experiment\ 2:}
For mean of the reward, we consider the same form as given in Experiment 1. Suppose that $S$ is independent of $\boldsymbol{X}$, following a Bernoulli distribution with $p=0.5$. 



\noindent\textbf{Experiment\ 3:}
We consider mean of the reward taking the form:
\begin{equation*}
\begin{aligned}
T(\boldsymbol{X}, S,A)=10+\left(0.1X_1^2-X_2-10SI(A=1)\right)A.
\end{aligned}
\end{equation*}
Assume that $S$ takes values in $\{-1, 0, 1\}$ with probabilities 0.25, 0.5, and 0.25, respectively.


\noindent\textbf{Experiment\ 4:}
We consider mean of the reward taking the form:
 \begin{equation*}
\begin{aligned}
T(\boldsymbol{X}, S,A)=10+ X_1+X_2+0.25 X_3+\left(X_1+X_2+10(S-1)^2\right)A.
\end{aligned}
\end{equation*}
Assume that $S$ takes values in $\{-1, 0, 1\}$ with probabilities 0.25, 0.5, and 0.25, respectively.

We apply a nonlinear proxy for all experiments. We consider a linear decision function for Experiments 1 and 2, and a nonlinear decision function with Gaussian kernels for Experiments 3 and 4. For the nonlinear decision function, the bandwidth parameter $\sigma$ and regularization parameter $\lambda$ are selected through a twofold cross-validation procedure, where a predefined finite set of $(\lambda, \sigma)$ is searched to find the pair that maximizes the empirical average of $\mathcal{V}(\mathcal{D})$ in Equation (\ref{e1}). For tuning parameter selection and policy training, we generate samples of size $n=200$ and $500$. The value function for the testing dataset, with a sample size of $500$, is estimated based on the empirical average of rewards obtained by executing the estimated DPA-ITR on the testing dataset. To assess performance, these experiments are repeated 200 times.  We will characterize the performance of DPA ITR using unfairness measure (UFM). The UFM is calculated as the difference value between the highest and lowest rates of accepting treatment within various sensitive groups.

\subsection{Results and conclusion}
\begin{figure}[h]
	\centering
 	\begin{minipage}{0.45\linewidth}
		\centering
		\includegraphics[width=1\linewidth]{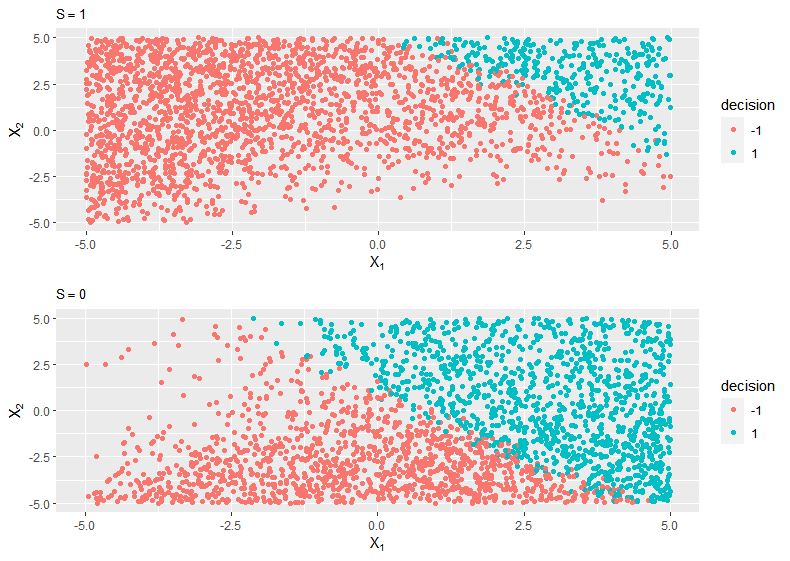}
		\subcaption{The OWL decision plot with.}
		\label{f5}
	\end{minipage}
        \hfill
        \begin{minipage}{0.45\linewidth}
		\centering
		\includegraphics[width=1\linewidth]{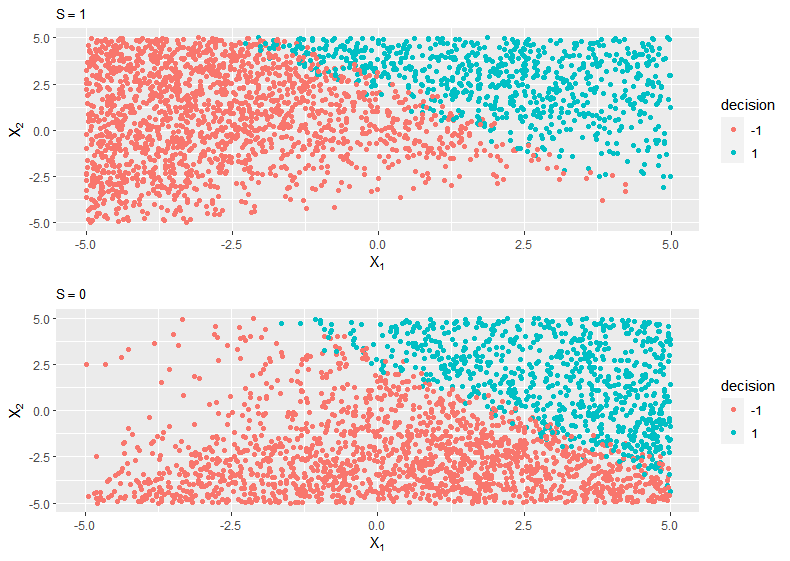}
	\subcaption{The DPA-ITR decision plot.}
		\label{f4}
	\end{minipage}
 	\begin{minipage}{0.45\linewidth}
		\centering
		\includegraphics[width=1\linewidth]{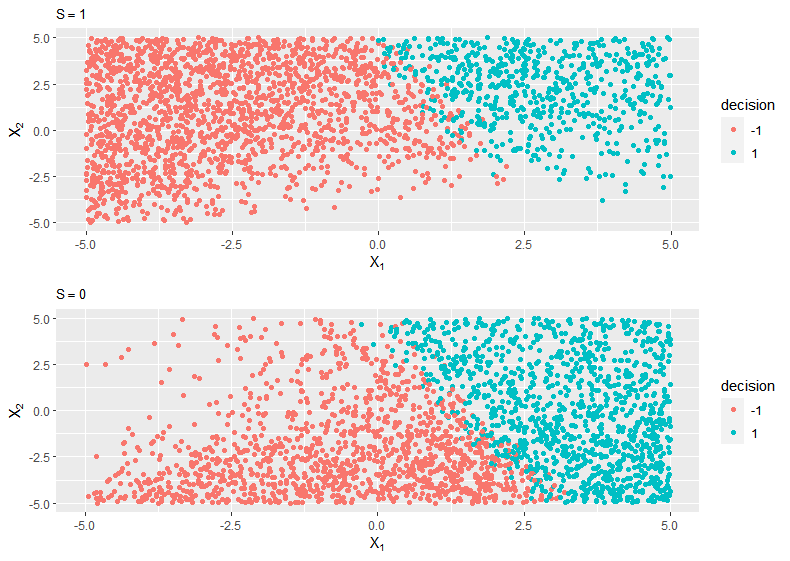}
		\subcaption{The naive method decision plot.}
		\label{f6}
 	\end{minipage}
 	\caption{The decision plot in Experiment 1 with $p=3$ $n=500$; the naive method  means estimating the ITR without considering the sensitive attribute $S$ via OWL; The UFM in Figure \ref{f5}, \ref{f6},\ref{f4} are $0.178$, $0.109$, $0.011$, respectively.}
    \label{f444}
\end{figure}

Figure \ref{f444} provides a comparison of decision plots for the OWL policy, the naive method, and DPA-ITR.  As depicted in Figures \ref{f4} and \ref{f5}, the DPA-ITR consistently outperforms the OWL approach in terms of the UFM, with decision points (red points and green points) appropriately aligned. This improvement in UFM can be attributed to the inherent bias present in the OWL policy, as evidenced by an unfairness measure of 0.178.  Figures \ref{f6} further illustrates that when the sensitive attribute is excluded from the decision function, achieving fairness remains difficult due to the intrinsic correlation between $\boldsymbol{X}$ and $S$.

\begin{table}
\caption{ {The table displays the estimated ${\widehat{\omega}}(\widehat{f})$ under different Experiments with $n=500$, where ${\widehat{\omega}}(\cdot)$ is derived using the the testing dataset and $\widehat{f}$ is obtained from the training dataset.}}
\centering
{\begin{tabular}{cccccccccc}
\toprule
   ${\widehat{\omega}}(\widehat{f})$ &{$c$} &0.02&0.04&0.06&0.08&0.10&0.12&0.14&0.16\\
       \hline
   \multirow{2}{*}{Experiment 1}&   $p=3$ &   $0.023$& $0.042$ &$0.062$& $0.081$&$ 0.099$ &$0.115$& $0.125$ &$0.131$\\
   &  $p=50$ &$0.023$  &$0.042$ &$ 0.061 $& $0.078  $ & $0.096 $ &  $0.109 $ & $0.118 $& $ 0.126 $\\
     \hline
   \multirow{2}{*}{Experiment 2}&   $ p=3  $&   $ 0.017 $&   $0.036 $&  $ 0.056  $& $ 0.075 $&  $ 0.094  $&  $0.109 $& $  0.119 $&  $ 0.124 $\\
   &  $p=50 $ & $0.018  $ & $0.038  $& $ 0.057 $ & $0.077  $ & $0.094  $& $0.109 $ & $0.1193  $ & $0.123 $\\
  \hline
    \multirow{2}{*}{Experiment 3}&   $ p=3  $& $0.020 $&  $ 0.041 $&  $ 0.061 $  &  $0.080 $ & $0.093  $& $ 0.100 $&   $ 0.101 $&  $  0.101$\\
   &  $ p=50 $ & $0.017$ & $0.036 $&  $0.053  $& $0.070  $& $0.081 $&  $0.085 $ & $0.085 $& $0.085 $\\
    \hline
    \multirow{2}{*}{Experiment 4}&   $ p=3 $ & $0.020 $ &  $0.039  $&  $0.056 $& $ 0.071  $& $0.081 $& $0.090 $ & $0.095  $& $0.098 $\\
   &   $p=50 $ & $0.017 $& $0.026  $& $0 .035 $& $0.043 $& $0.048 $&  $0.048  $& $0.048  $& $0.048   $\\
     \bottomrule
\end{tabular}}

\end{table}

\begin{figure}[h]
	\centering
 	\begin{minipage}{0.45\linewidth}
		\centering
	\includegraphics[width=1\linewidth]{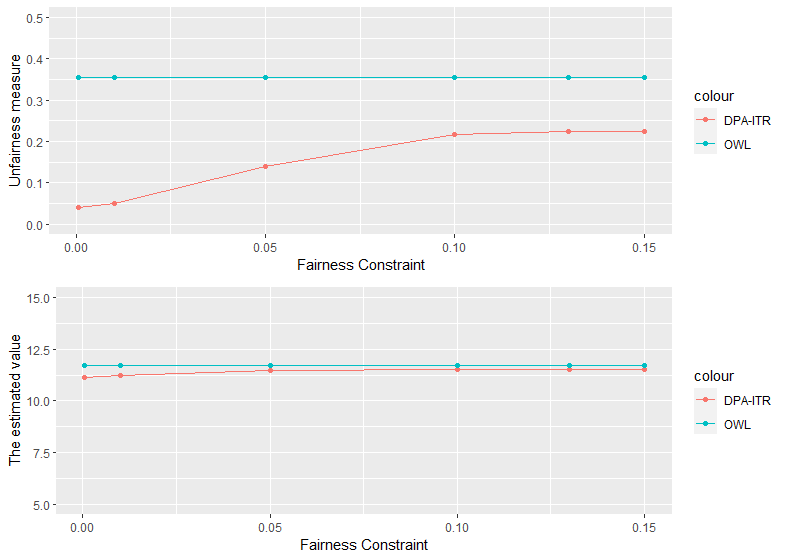}
        \subcaption{Experiment 1 with $p=3$}
		\label{f8}
	\end{minipage}
        \hfill
  	\begin{minipage}{0.45\linewidth}
		\centering
	\includegraphics[width=1\linewidth]{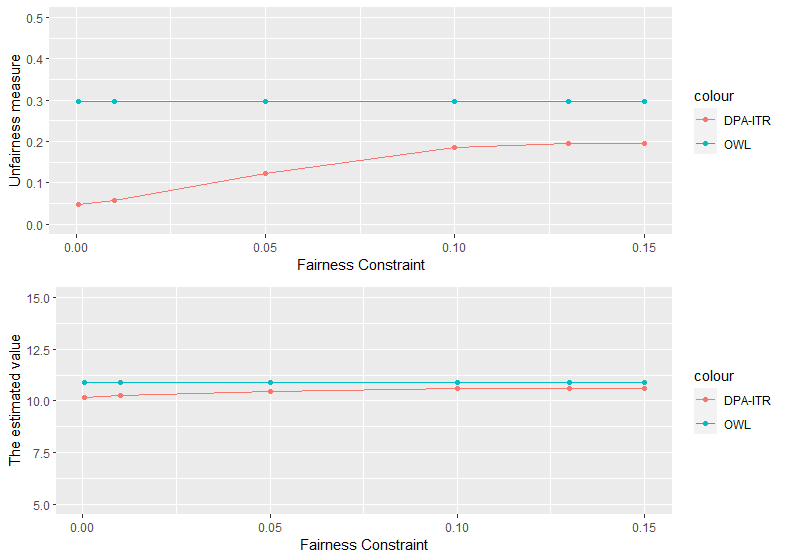}
        \subcaption{Experiment 1 with $p=50$}
		\label{f9}
	\end{minipage}
 \hfill
 	\begin{minipage}{0.45\linewidth}
		\centering
	\includegraphics[width=1\linewidth]{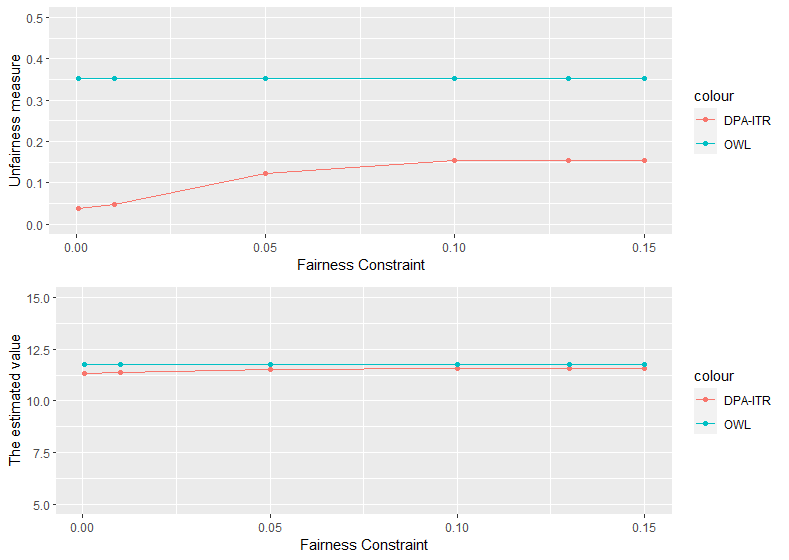}
        \subcaption{ Experiment 2 with $p=3$}
		\label{f10}
	\end{minipage}
        \hfill
  	\begin{minipage}{0.45\linewidth}
		\centering
	\includegraphics[width=1\linewidth]{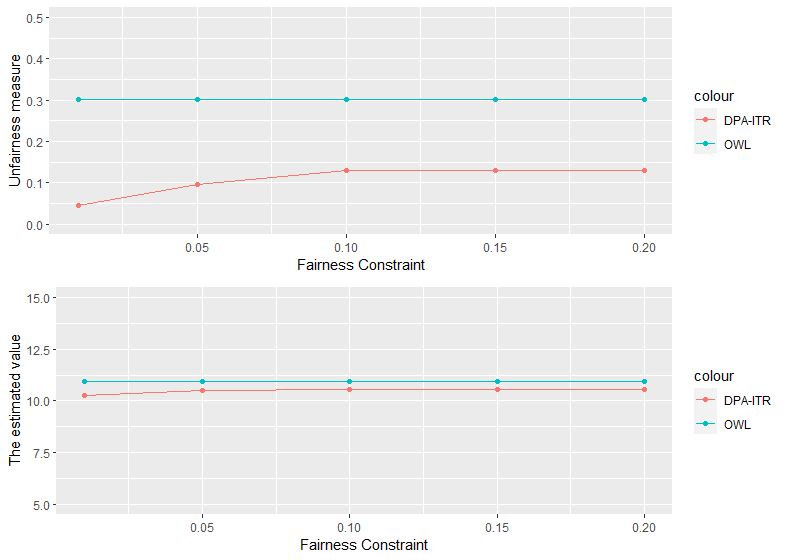}
        \subcaption{Experiment 2 with $p=50$}
		\label{f11}
	\end{minipage}
         \caption{The UFM and the estimated value plot  for training sample size $n=500$.}
         \label{fig4}
\end{figure}

\begin{figure}[h]
	\centering
  	\begin{minipage}{0.45\linewidth}
		\centering
	\includegraphics[width=1\linewidth]{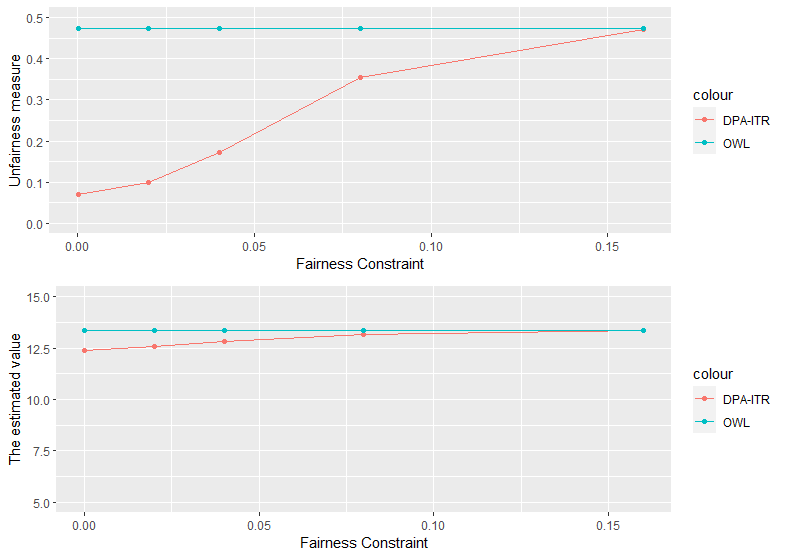}
        \subcaption{Experiment 3 with $p=3$}
		\label{f12}
	\end{minipage}
        \hfill
  	\begin{minipage}{0.45\linewidth}
		\centering
	\includegraphics[width=1\linewidth]{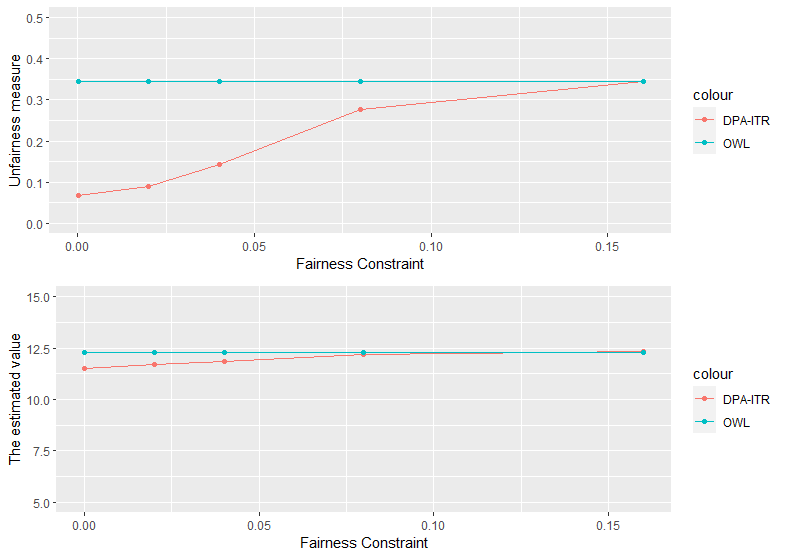}
        \subcaption{Experiment 3 with $p=50$}
		\label{f13}
	\end{minipage}
    \hfill
 	\begin{minipage}{0.45\linewidth}
		\centering
	\includegraphics[width=1\linewidth]{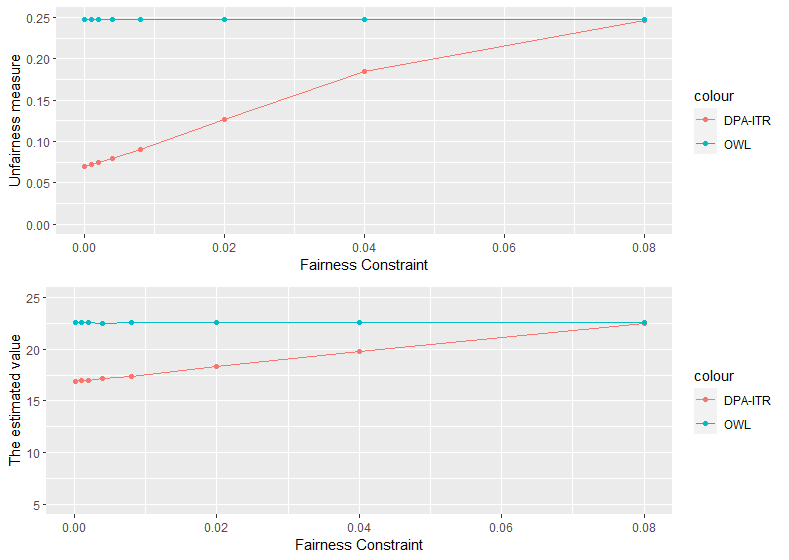}
        \subcaption{Experiment 4 with $p=3$}
		\label{f14}
	\end{minipage}
        \hfill
  	\begin{minipage}{0.45\linewidth}
		\centering
	\includegraphics[width=1\linewidth]{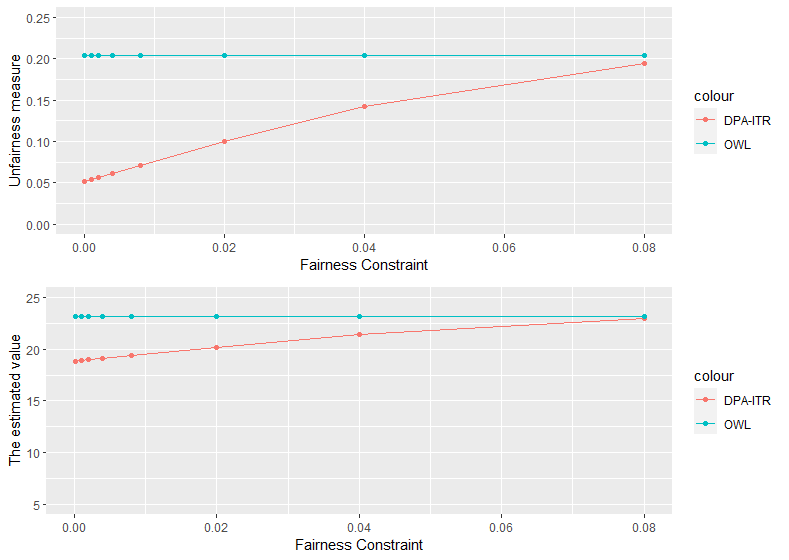}
        \subcaption{Experiment 4 with $p=50$}
		\label{f15}
	\end{minipage}
         \caption{The UFM and the estimated value plot  for training sample size $n=500$.}
         \label{f555}
\end{figure}

 The estimated $\widehat{\omega}(\widehat{f})$ in Table 1 shows close agreement with the corresponding values of $c$, providing validation for the convex property shown in Theorem 1. Figures \ref{f444}-\ref{f555} plot the UFM across different experiments.

\begin{table}
\caption{ { The table displays the estimate ${\widehat{\omega}}(\widehat{f})$ under different Experiments with $n=200$.}}
\centering
{\begin{tabular}{cccccccccc}
\toprule
   ${\widehat{\omega}}(\widehat{f})$ &{$c$} &0.02&0.04&0.06&0.08&0.10&0.12&0.14&0.16\\
   \hline
   \multirow{2}{*}{Experiment 1}&   $p$=3 &   0.024& 0.041 &0.058& 0.074& 0.088 &0.099& 0.107 &0.114\\
   &  $p$=50 &0.023  &0.042 & 0.061  &0.078  &0.096 & 0.109 &0.118 & 0.126\\
   \hline
   \multirow{2}{*}{Experiment 2}&   $p$=3 &   0.026&  0.046&  0.038 & 0.074&  0.090 & 0.127&  0.123&  0.137\\
   &  $p$=50 &0.028  &0.046 & 0.071 &0.084  &0.099 &0.129  &0.137  &0.152\\
   \hline
    \multirow{2}{*}{Experiment 3}&   $p$=3 &0.024&  0.043&  0.061 & 0.080 & 0.096 & 0.109&   0.118&   0.122\\
   &  $p$=50 &0.014 &0.029 &0.041 &0.051& 0.057 &0.058&0.058 &0.058\\
   \hline
    \multirow{2}{*}{Experiment 4}&   $p$=3 &0.019&  0.033&  0.042  & 0.047&0.050&0.051&0.051&0.051\\
   &  $p$=50 &0.014 &0.022 &0.028&0.032&0.034& 0.034 &0.034 &0.034\\
     \bottomrule
\end{tabular}}

\end{table}

\begin{figure}[h]
	\centering
 	\begin{minipage}{0.45\linewidth}
		\centering
	\includegraphics[width=1\linewidth]{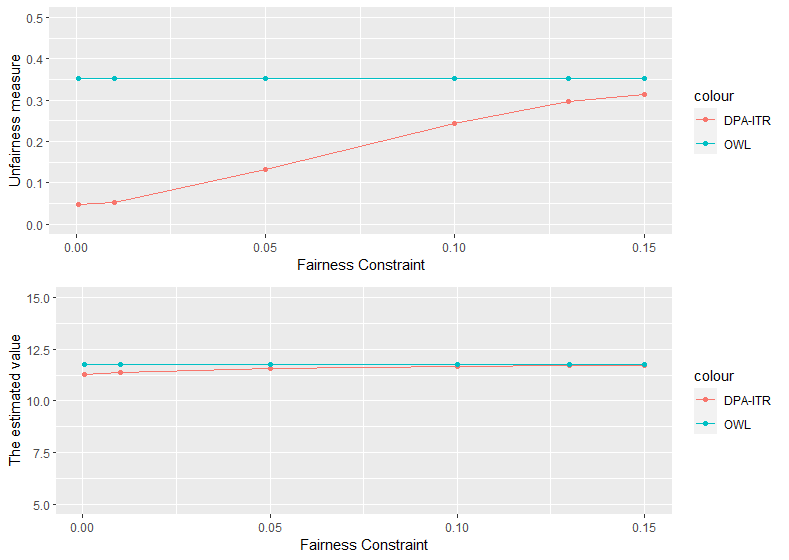}
        \subcaption{Experiment 1 with $p=3$}
		\label{f17}
	\end{minipage}
        \hfill
  	\begin{minipage}{0.45\linewidth}
		\centering
	\includegraphics[width=1\linewidth]{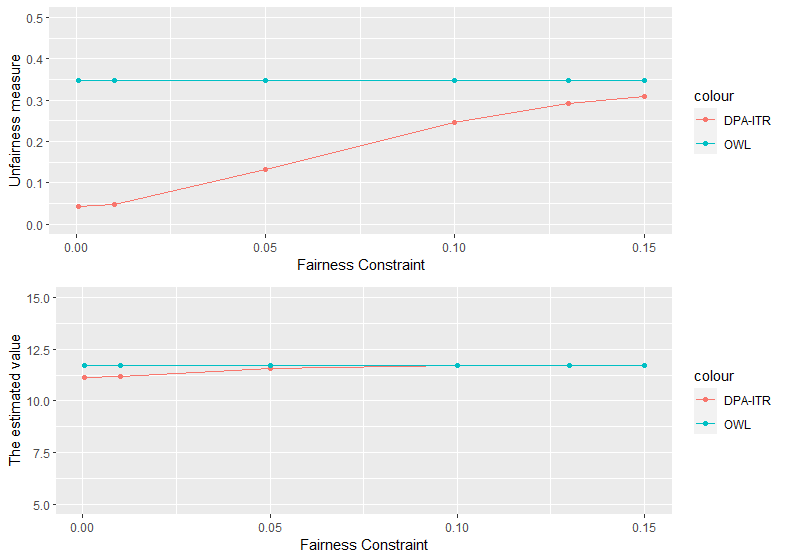}
        \subcaption{Experiment 1 with $p=50$}
		\label{f18}
	\end{minipage}
 	\begin{minipage}{0.45\linewidth}
		\centering
	\includegraphics[width=1\linewidth]{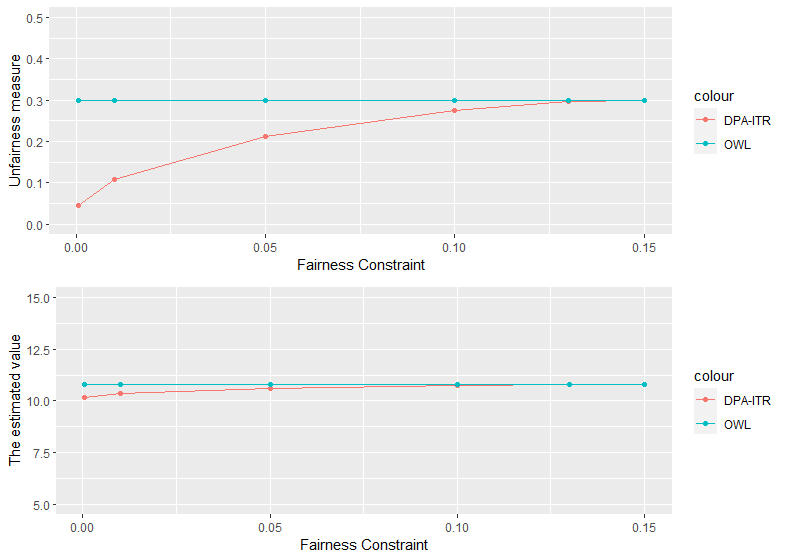}
        \subcaption{ Experiment 2 with $p=3$}
		\label{f19}
	\end{minipage}
        \hfill
  	\begin{minipage}{0.45\linewidth}
		\centering
	\includegraphics[width=1\linewidth]{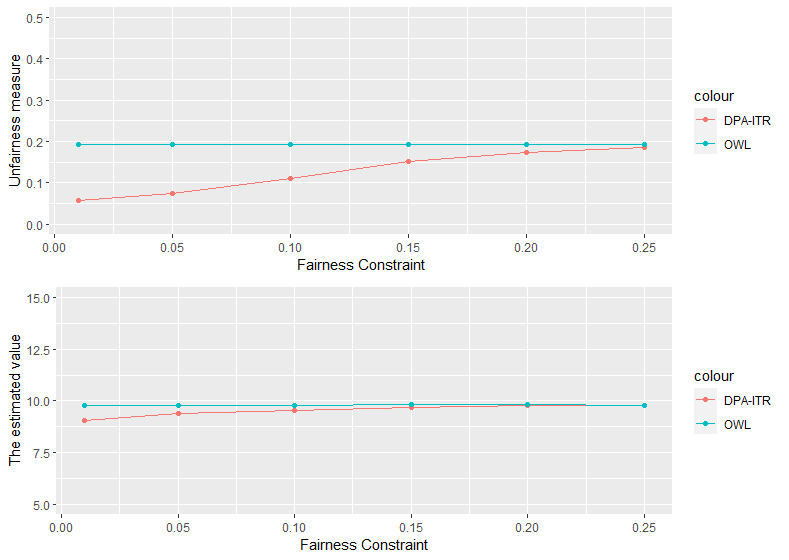}
        \subcaption{Experiment 2 with $p=50$}
		\label{f20}
	\end{minipage}
         \caption{The UFM and the estimated value plot for training sample size $n=200$.}
\end{figure}
\begin{figure}[h]
	\centering
        \begin{minipage}{0.45\linewidth}
		\centering
	\includegraphics[width=1\linewidth]{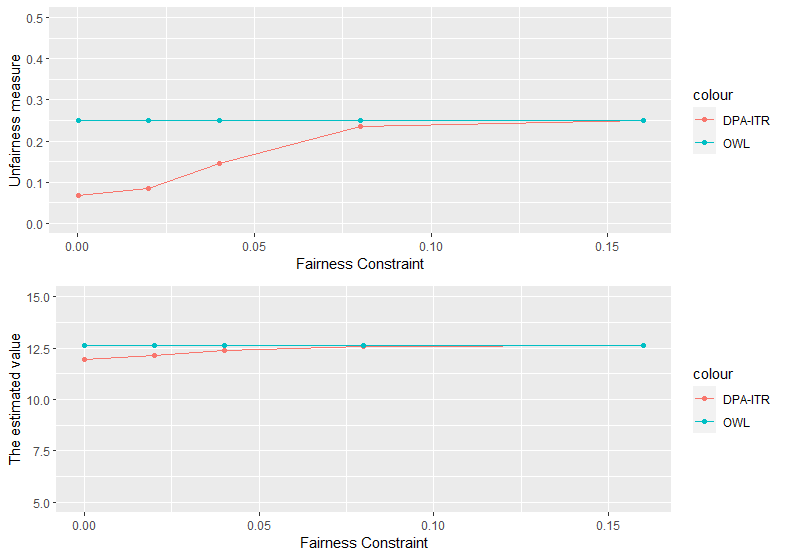}
        \subcaption{Experiment 3 with $p=3$}
		\label{f21}
	\end{minipage}
         \hfill
  	\begin{minipage}{0.45\linewidth}
		\centering
	\includegraphics[width=1\linewidth]{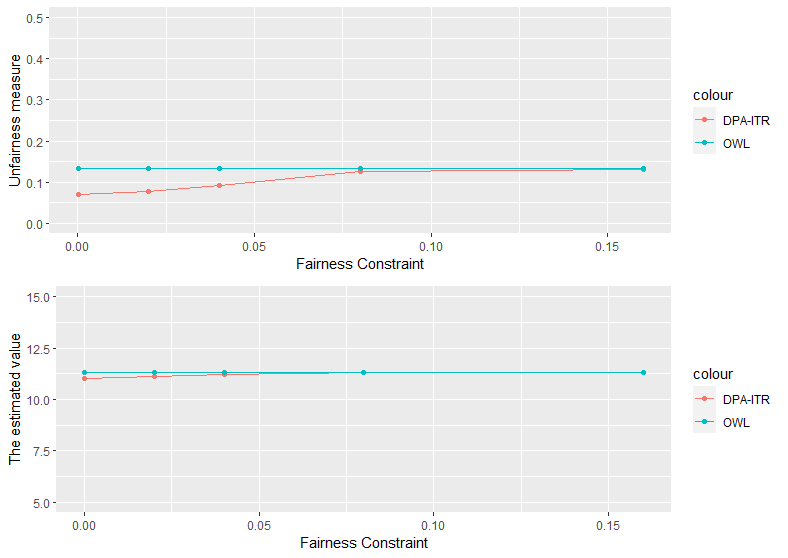}
        \subcaption{Experiment 3 with $p=50$}
		\label{f22}
	\end{minipage}
  \hfill
 	\begin{minipage}{0.45\linewidth}
		\centering
	\includegraphics[width=1\linewidth]{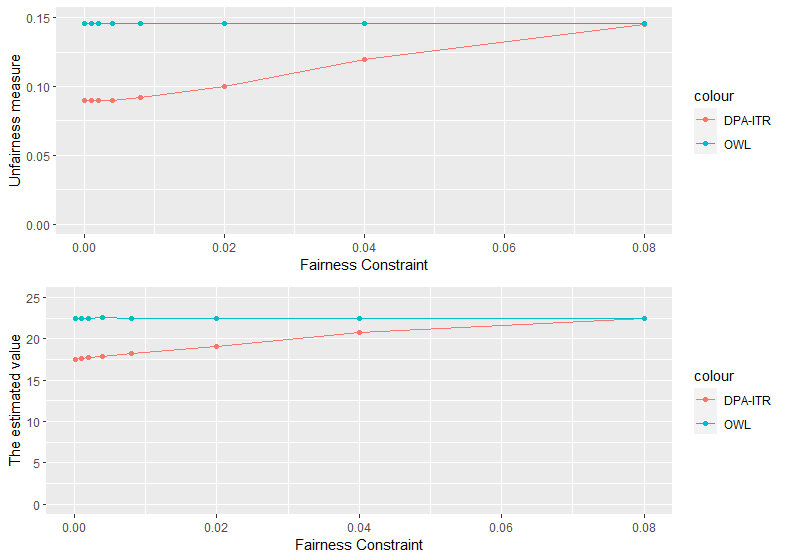}
        \subcaption{ Experiment 4 with $p=3$}
		\label{f23}
	\end{minipage}
        \hfill
  	\begin{minipage}{0.45\linewidth}
		\centering
	\includegraphics[width=1\linewidth]{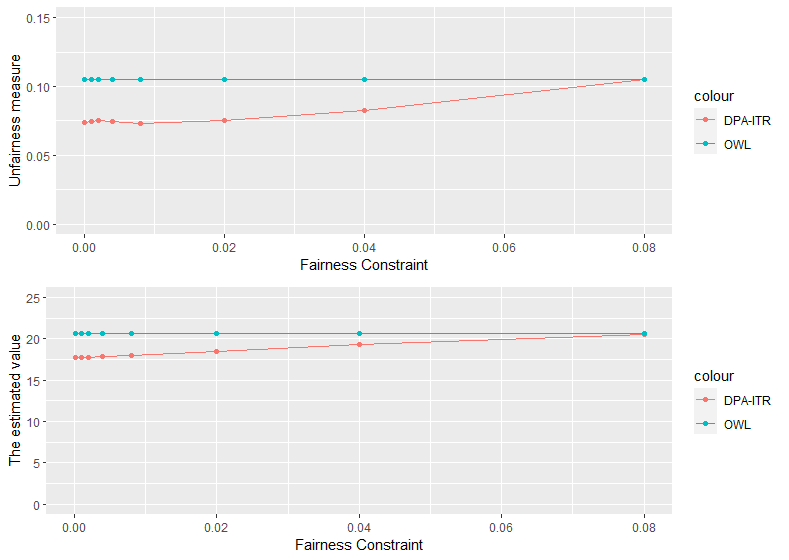}
        \subcaption{Experiment 4 with $p=50$}
		\label{f24}
	\end{minipage}
         \caption{The UFM and the estimated value plot for training sample size $n=200$.}
         \label{f777}
\end{figure}

Figure \ref{fig4}-\ref{f777} demonstrates that reducing the fairness constraint $c$ leads to a decrease in UFM as well as a decrease in policy value, indicating a tradeoff between fairness and policy value. In Experiments 3 and 4, where a nonlinear reward model is considered, the DPA-ITR consistently demonstrates high effectiveness, as shown in Figure \ref{f12}-\ref{f15}. The DPA-ITR successfully achieves a UFM of less than 0.10, illustrating its capability to effectively manage fairness even in complex scenarios.

\subsection{Extension to empirical selection for parameter $c$}
\begin{figure}[h]
	\centering
 	\begin{minipage}{1\linewidth}
		\centering
	\includegraphics[width=0.45\linewidth]{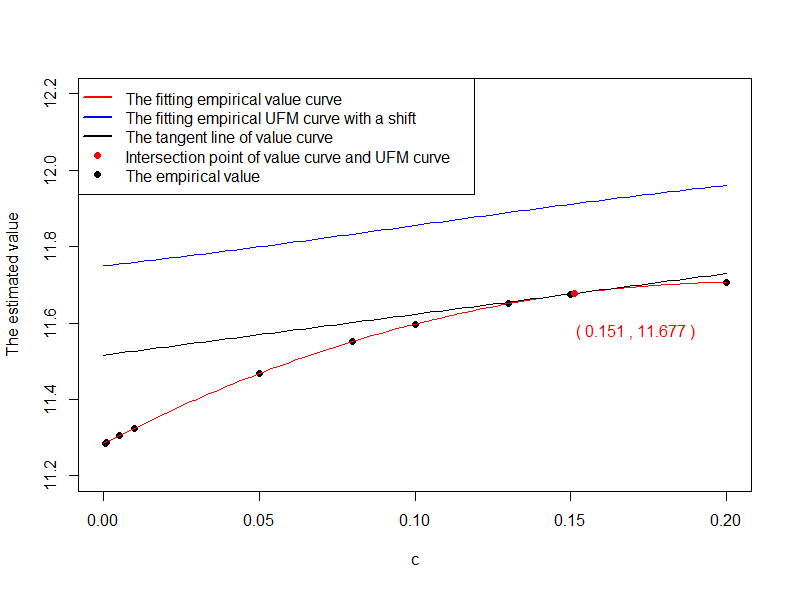}
        \caption{The empirical UFM and value plot in Experiment 1 with $p=3$.}
		\label{f16}
	\end{minipage}
\end{figure}

The above experiments show that decreasing the fairness constraint $c$ leads to an improvement in fairness but comes at the cost of reducing the policy value. Therefore, we propose two methods of selecting the fairness constraint $c$ through empirical experiments.

Define $V_0(c) = \mathcal{V}(\mathcal{D}(\widehat{f}_n))$ as the value achieved under the fairness constraint, and $U_0(c)$ as the unfairness measure of the policy $\mathcal{D}(\widehat{f}_n)$ under fairness constraint $c$.
Using Experiment 1 as an example, we could get empirical value of ${V}_0(c)$ and ${U}_0(c)$ for  different constraint $c$. Then, we utilize  polynomial functions $V(c)$ and $U(c)$ to approximate $V_0(c)$ and $U_0(c)$ based on these empirical values.  Figure \ref{f16} illustrates that  $U(c)$ decreases at a faster rate compared to  $V(c)$ as $c$ decreases. This implies that a small sacrifice in policy value can result in a substantial improvement in the fairness measure. To determine an optimal point, denoted as $c_0$, we identify the minimum value of $c$ that satisfies the condition $U'(c) - V'(c) > 0$, where $U'(c)$ and $V'(c)$ represent the derivatives of the functions $U(c)$ and $V(c)$, respectively. We refer to this point $c$ as the most cost-effective point. Figure \ref{f16} demonstrates the identification of the most cost-effective point at $0.151$ by solving the inequality $U'(c) - V'(c) > 0$.

{
The ``four-fifths rule"  \citep{biddle2017adverse} is a widely used criterion for assessing discrimination. According to this rule, a policy is considered fair if the selection rate for any race, sex, or ethnic group is at least four-fifths (80\%) of the rate for the group with the highest selection rate. We can also empirically select $c$ to satisfy the ``four-fifths rule".}

\section{Application}


We use the proposed method to analyze data sourced from the Next 36 program, an entrepreneurship program for undergraduate students in North America \citep{lyons2017impact}. The program started accepting applicants in 2011 and conducts one session per year. The objective of this program is to foster the next generation of innovators by providing coursework, mentorship, financial capital, and access to the entrepreneurship community. Participants gain the opportunity to explore the establishment of their own ventures. Our goal is to develop a policy for allocating undergraduate students to the entrepreneurial program based on their individual characteristics while ensuring fairness with respect to gender. The dataset used in this analysis is derived from the Next 36 program sessions conducted between 2011 and 2015 and consists of 335 observations, including 179 accepted applicants $(A=1)$ and 156 applicants who were not accepted $(A=-1)$. This quasi-experiment aims to optimize subsequent entrepreneurial activity, which is measured using a binary variable $(R=1$ or $-1$) indicating whether participants were involved in a startup after the program ended.

\begin{table}
 \caption{  The performance table with $5$-fold cross-validation; ``Naive'' means learning the ITR without considering the sensitive attribute $S$ via OWL;  we used nonlinear decision function with linear  fairness proxy and nonlinear  fairness proxy for estimating DPA-ITR.}
 \centering
{\begin{tabular}{cccc}
  \toprule
    $c=0.00001$ & Method  & The estimated value & UFM \\
      \midrule
    \multirow{4}{*}{Case1 } & OWL  & 0.271 &0.137   \\
   & Naive  & 0.203& 0.091\\
      & Linear  Fairness Proxy  & 0.267& 0.078\\
      & Nonlinear  Fairness Proxy  & 0.220& \textbf{0.066}\\
     \midrule
    \multirow{4}{*}{Case2 } & OWL  & 0.297 &0.079 \\
   & Naive  & 0.303& 0.090\\
      & Linear Fairness Proxy  & 0.297& 0.071\\
      & Nonlinear Fairness  Proxy  & 0.262& \textbf{0.052}\\
  \bottomrule
\end{tabular}}
 \label{table2}

\end{table}



We consider two cases based on different covariates. In Case 1, we include the average score assigned by the interviewer and school rank as covariates. In Case 2, we expand the covariates to include the average score assigned by the interviewer, school rank, race (white or non-white), years to graduation, prior entrepreneurship activities, major, and other relevant factors. In both cases, we consider gender as the sensitive attribute $S$. To learn the nonlinear ITR, we employ a Gaussian kernel. We compare our proposed approach to the OWL method introduced by \cite{zhao2012estimating} and a naive method where we disregard the sensitive attribute during the learning process, thereby completely ignoring gender fairness. Following \cite{lyons2017impact} and \cite{viviano2023fair}, we estimate the propensity score functions using penalized logistic regression.

Table \ref{table2} demonstrates that our proposed method can learn the fairest policy when measuring unfairness as the difference between admission rates for the female and male groups.
It can be seen from Table \ref{table2} that the use of nonlinear fairness proxy is able to significantly reduce the unfairness.


\section{Concluding remarks}
We have proposed a novel methodology for finding a fair policy that fulfills the demographic parity prerequisite by utilizing linear and nonlinear fairness proxies. To address the challenge associated with optimizing a nonlinear value function with fairness constraints, we have formulated the problem as convex quadratic programming. We have established the theoretical properties including the consistency  of the proposed estimator and the risk bound. The good performance of the proposed method has been demonstrated through comprehensive simulation studies and application to data from the Next 36 program.

In future research endeavors, our study paves the way for exploring the development of an optimal ITR that operates completely independent of the sensitive attribute, moving beyond the reliance on fair proxies. It is also interesting to estimate the demographic parity-aware ITR by adding  constraints for fairness control.
Furthermore, we recognize  the delicate balance required to uphold fairness while optimizing value. This complex issue demands thorough exploration and consideration as we advance.

\section*{Supplementary material}
The supplementary material includes all theoretical proofs.

\bibliographystyle{chicago}
\bibliography{paper-ref}

\end{document}